\documentclass[a4paper,fleqn]{cas-sc}
\usepackage{lineno,hyperref}
\modulolinenumbers[5]
\usepackage{framed,multirow}
\usepackage[nottoc]{tocbibind}

\newcommand{\tabincell}[2]{\begin{tabular}{@{}#1@{}}#2\end{tabular}}
\usepackage{float}
\usepackage[authoryear]{natbib}
\def\tsc#1{\csdef{#1}{\textsc{\lowercase{#1}}\xspace}}
\tsc{WGM}
\tsc{QE}
\tsc{EP}
\tsc{PMS}
\tsc{BEC}
\tsc{DE}

\begin{document}
\let\WriteBookmarks\relax
\def\floatpagepagefraction{1}
\def\textpagefraction{.001}
\shorttitle{Accurate and Robust Pulmonary Nodule Detection}
\shortauthors{CV Liu et~al.}
\title [mode = title]{Accurate and Robust Pulmonary Nodule Detection by 3D Feature Pyramid Network with Self-supervised Feature Learning} \date{July 24, 2019}
\author[1]{Jingya Liu}
\author[2,3]{Liangliang Cao}
\author[4]{Oguz Akin}
\author[1]{Yingli Tian\corref{cor1}}
\cortext[cor1]{Corresponding author}
\ead{ytian@ccny.cuny.edu}

\address[1]{The City College of New York, New York, NY 10031, USA}
\address[2]{UMass CICS, Amherst, MA 01002, USA}
\address[3]{Google AI, New York, NY 10011, USA}
\address[4]{Memorial Sloan Kettering Cancer Center, New York 10065, USA}

\begin{abstract}

Accurate detection of pulmonary nodules with high sensitivity and specificity is essential for automatic lung cancer diagnosis from CT scans. 
Although many deep learning-based algorithms make great progress for improving the accuracy of nodule detection, the high false positive rate is still a challenging problem which limits the automatic diagnosis in routine clinical practice. Moreover, the CT scans collected from multiple manufacturers may affect the robustness of Computer-aided diagnosis (CAD) due to the differences in intensity scales and machine noises. 
In this paper, we propose a novel self-supervised learning assisted pulmonary nodule detection framework based on a 3D Feature Pyramid Network (\textit{3DFPN}) to improve the sensitivity of nodule detection by employing multi-scale features to increase the resolution of nodules, as well as a parallel top-down path to transit the high-level semantic features to complement low-level general features. 
Furthermore, a High Sensitivity and Specificity (\textit{HS$^2$}) network is introduced to eliminate the false positive nodule candidates by tracking the appearance changes in continuous CT slices of each nodule candidate on  Location History Images (LHI). 
In addition, in order to improve the performance consistency of the proposed framework across data captured by different CT scanners without using additional annotations, an effective self-supervised learning schema is applied to learn spatiotemporal features of CT scans from large-scale unlabeled data. 
The performance and robustness of our method are evaluated on several publicly available datasets including LUNA16, SPIE-AAPM, LungTIME, and HMS. 
The proposed framework is able to accurately detect pulmonary nodules with high sensitivity and specificity and achieves $90.6\%$ sensitivity with $1/8$ false positive per scan which outperforms the state-of-the-art results $15.8\%$ on LUNA16 dataset. The significant performance improvements on the three addition datasets ($7.4\%$ on SPIE-AAPM, $13.5\%$ on LungTIME, and $8.9\%$ on HMS respectively) compared with the results without using pre-trained model demonstrate the effectiveness of the self-supervised learning schema to handle data captured by different brands of CT scanners. 

\end{abstract}

\maketitle

\section{Introduction}
Lung cancer is one of the leading cancer killers around the world which makes the study of lung cancer diagnosis eminently crucial \citep{litjens2017survey,ohno2017standard}. Computer-aided diagnosis (CAD) systems provide assistance for radiologists to accelerate the diagnosing process~\citep{hosny2018artificial,pesce2019learning}. Many efforts \citep{huang2017lung,ramachandran2018using,wang2017central} have been made for lung nodule detection on CT scans by applying the recent powerful deep detection models in computer vision. These efforts made good progress to accurately detect pulmonary nodules in curated datasets, but underperform in different datasets and in real clinical settings. In addition, the false positive rate is still very high which limits the real application in routine clinical practice. For example, most of the previous work \citep{Ding2017Accurate,dou2017multilevel,khosravan2018s4nd,wang2018automated,zhu2018deeplung} obtained less than $75\%$ sensitivity with $1/8$ false positives per scan. To obtain sensitivity scores higher than $95\%$, these models would bear about eight false positives per scan. Furthermore, to maintain consistent performance across the data captured by different brands of CT scanners, there is a need to continuously adjust and re-train the networks which generally require a vast amount of training data to be labeled. However, manual labeling of large numbers of CT scans is tedious, attention-demanding, and very time-consuming. It also requires knowledge and skills of a particular subspecialty of radiology. 

There are three main challenges for accurate and robust lung nodule detection: 
1) The robustness of existing deep learning-based lung nodule detection methods is limited by the noise among CT scans from different equipment manufacturers. Various CT image texture reconstruction methods and image acquisition techniques are applied to CT scans from different manufacturers during the data collection process. For the data-dependent neural network based framework, artifacts such as intensity, machine setting, machine noise, and image protocol could cause systematical differences and affect the robustness of the nodule detection framework. 
2) Some normal tissues have similar morphological appearances as nodules in CT images which cause high false positives by wrongly detecting these tissues as nodules. The approach for differentiating between the tissue and nodule is very crucial to reduce false positives for automatic lung nodule detection scheme. 
3) The high discrepancy of the volume between nodules and the whole CT scan may cause missing detection of real nodules. For example, the volume of a nodule with 10 mm size in the diameter only occupied $0.059\%$ of the volume of whole CT scan (in average $213\times293$ pixels with $281$ slices). Furthermore, the size of pulmonary nodules can vary by as much as 10 times. For example, nodule sizes can range from 3 mm to 30 mm (in diameter) in LUNA16 dataset. 

Therefore, it is essential to design methods which can detect on CT scans with various machines settings and intensity scales, for the small volume nodules from large volume CT scans, as well as to differentiate nodules from tissues with similar appearances in CT images. 

\begin{figure}
    \centering
    \includegraphics[width=0.9\textwidth]{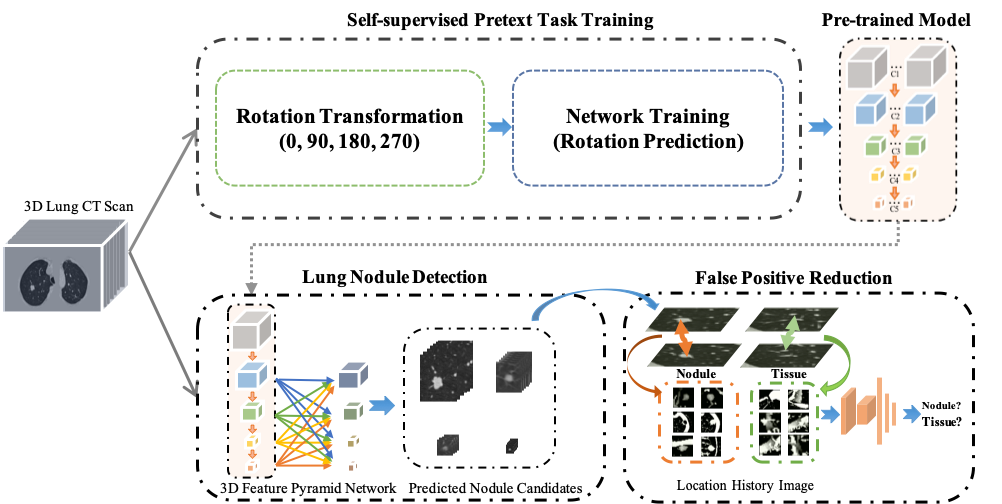}
    \caption{\textbf{The proposed self-supervised learning assisted \textit{3DFPN-HS$^2$} framework of high sensitivity and specificity lung nodule detection by combining a 3D Feature Pyramid ConvNet (\textit{3DFPN}) with a self-supervised pre-trained model and an \textit{HS$^2$} false positive reduction network.} 1) To improve the robustness across different datasets, a self-supervised learning-based pre-trained \textit{ResNet-18} model is applied to the nodule detection network. The pre-trained model is obtained by a self-supervised pretext task with a rotation-transformation network for CT scan geometric transformation at 0, 90, 180, 270 degrees followed by a network training procedure to predict the rotated angles of input 3D CT scans.
    2) A whole CT scan is fed into the \textit{3DFPN} to predict nodule candidates. The backbone network (\textit{ResNet-18}) of \textit{3DFPN} is trained as the pre-trained model and then is fine-tuned on the dataset for lung nodule detection.
    3) For the detected nodule candidates, the \textit{HS$^2$} network eliminates the falsely predicted normal tissues based on the location variance within continuous CT slices on LHI images. 
    The detailed structure of self-supervised pretext task training is shown in Figure~\ref{fig:self-supervised}, the proposed \textit{3DFPN} network can be found in Figure~\ref{fig:3DFPN}, and LHI is illustrated in Figure~\ref{fig:candiates}. 
    } \label{fig:system}
\end{figure}

To address the above three challenges, this paper attempts to integrate the most recent progress in computer vision as well as the domain expert knowledge in medical imaging. As shown in Figure~\ref{fig:system}, the proposed framework contains three main components. 
1) Inspired by the rotation ConvNets~\citep{gidaris2018unsupervised,jing2018self}, to improve the robustness of the proposed nodule detection framework across data captured by different CT scanners without using additional annotations, the self-supervised learning method with 3D CT scan rotation prediction network is proposed to obtain a pre-trained model to extract spatiotemporal features and leverage the differences among CT scans from various manufacturers to guarantee the robustness of the proposed detection framework. 
2) Motivated by the state-of-the-art image detector using 2D Feature Pyramid Network (\textit{FPN}) \citep{lin2017feature}, a 3D high sensitivity feature pyramid network (\textit{3DFPN}) is developed to detect nodules of different sizes by appending the low-level high-resolution features to the high-level strong semantic features and a multi-scale feature prediction to ensure the wide-scale nodule detection. 
3) A false positive reduction network is proposed to achieve high specificity while keeping high sensitivity. By carefully analyzing the difference between nodules and the tissues which are wrongly detected as nodule candidates, although they look similar on single CT slice, we observe that the distributions of their spatial variances among continuous CT slices are very different. This unique insight motivates us to design a novel refinement network based on the location history image in the continuous CT slices which significantly eliminates the false positive numbers. 

This paper presents an extension to~\cite{jliu2019miccai}, the main contributions are described as follows:
\begin{enumerate}
    \item We propose a novel and accurate two-stage framework \textit{3DFPN-HS$^2$} for pulmonary nodule detection by integrating a nodule detection model with a false positive reduction scheme to achieve both high sensitivity and specificity.
    \item A self-supervised learning-based pre-trained model is introduced to improve the robustness of the proposed \textit{3DFPN-HS$^2$} framework across CT scans captured from multiple manufacturers without requiring additional labels. The pre-trained model could be further applied to other related tasks with lung CT scans, such as pulmonary nodule classification, and etc., which save the effort for collecting large labeled dataset and also brings great benefits to the study on the small datasets.
    \item Our proposed framework achieves state-of-the-art performance on LUNA16 dataset at all the false positive levels with a significant improvement of the sensitivity, especially at the low false positive rates. It is robust to CT scans with multiple manufacturers on the different datasets, which makes the proposed work highly likely to be applied in clinical practice.
\end{enumerate}

The remainder of this paper is organized as follows. Section 2 introduces the related work on self-supervised feature learning, object detection, and lung nodule detection from CT scans. Section 3 describes the proposed method. In Section 4, a variety of implementation details, experimental results, and discussions are presented. Finally, section 5 summarizes the remarks of this paper.

\section{Related Work}
As ConvNets are data-driven computational mechanics, supervised frameworks generally require large-scale labeled data to achieve good performance and overcome overfitting. For lung nodule detection from CT scans, manual labeling of large numbers of CT scans is tedious, attention-demanding, and very time-consuming. It also requires knowledge and skills of a particular subspecialty of radiology. In addition, multiple expert radiologists are required to perform this task to address reader agreement and variability issues. Therefore, a self-supervised scheme may contribute to the applications without requiring additional data annotations. In computer vision research, several frameworks~\citep{doersch2015unsupervised,doersch2017multi,mundhenk2018improvements,jing2019self} have been proposed to learn image and video feature representation based on data transformation. \cite{lee2017unsupervised} adopt the learning task by verifying the correct order of the shuffled sequences from continuous video clips.~\cite{noroozi2016unsupervised} propose a puzzle-solving method to obtain spatial context information. \cite{gidaris2018unsupervised} learn image features by applying 2D ConvNets for image rotation prediction. However, previous methods are based on 2D which lack spatiotemporal information. Recently, \cite{jing2018self} propose a self-supervised learning model \textit{3DRotNet} for 3D sequences as input adopting a rotation prediction task to learn the rich spatiotemporal features. In this paper, by treating each CT scan as a video, we follow the framework proposed in~\cite{jing2018self} to obtain a pre-trained model for 3D CT scan feature representation extraction to improve the robustness of the lung nodule detection framework across different datasets.

Several deep learning-based frameworks for object detection are proposed to handle small and multi-scale objects~\citep{lin2017feature,jenuwine2018lung,liu2016ssd,singh2018analysis}. Single Shot multi-box Detector (\textit{SSD})~\citep{liu2016ssd} applies a pyramid feature hierarchy in the deep convolutional network to directly detect objects in multi-scales using feature maps from multiple layers by a single pass. However, \textit{SSD} cannot reuse the low-level feature maps which result in missing small objects. In order to detect small objects, Scale Normalization for Image Pyramids (\textit{SNIP})~\citep{singh2018analysis} selectively back-propagate the gradient of objects in different scales as a function of image scale. Although the performance of small object detection has been significantly improved, by applying the multiple images as the input, the computation cost could be very high. To date, a 2D Feature Pyramid Network (\textit{FPN})~\citep{lin2017feature} demonstrates the effectiveness for small object detection by extracting the multi-scale feature maps which contain general low-level features of objects at different scales. A top-down path is introduced to pass the global context information by lateral connections of the high-level features and low-level features. The computation cost of extracting features reduced by directly applying multi-scale feature maps. This \textit{FPN} framework can be applied to lung nodule detection on each 2D CT slices. However, without 3D temporal information among CT slices, high false positives are produced.

Many efforts have been made for lung nodule detection from CT scans. Compared to the traditional method with intensity, shape, texture features, and context features~\citep{hardie2008performance,matsumoto2006pulmonary,messay2010new,Murphy2009A,van2010comparing}, the deep learning-based methods show significant improvements on the performance~\citep{ciompi2015automatic,jacobs2014automatic,Setio2016Pulmonary,shin2016deep}. \cite{Ding2017Accurate} apply a two-stage framework of \textit{2D Faster R-CNN} nodule detector with false positive reduction classifier and obtain $89.1\%$ average sensitivity, while \cite{zhu2018deeplung} could achieve an average sensitivity of $84.2\%$ by a one-stage \textit{3D Faster R-CNN} detector to learn the rich nodule features combined with a dual-path network and a U-net like encoder-decoder structure without false positive reduction. Single-Shot Single-Scale Lung Nodule Detection (\textit{S4ND})~\citep{khosravan2018s4nd} introduce 3D dense connection and investigated a down-sampling method for small nodule detection. These frameworks employ only a single scale feature map causing limitations in detecting nodules with a wide size range.
\cite{gupta2018automatic} propose a multisize nodule detection method. The method first segments the lung boundary delineation to obtain the lung region and then applies three sub-algorithms to detect the candidates in three nodule size intervals. Although nodules with different sizes are treated separately, the rule-based thresholding and morphological algorithms limit the sensitivity to $85.6\%$ at 8 false positives per scan.
In addition, for false positive reduction,~\cite{dou2017multilevel,dou2017automated} adopt three different 3D ConvNet architectures trained for adapting the scales of nodules and manually setting the threshold to combine the weights. Multi-scale Gradual Integration Convolutional Neural Network (\textit{MGI-CNN})~\citep{kim2019multi} use the image pyramid network with a multi-stream feature integration for small nodule detection and false positive reduction. Nevertheless, computation costs for these frameworks are expensive due to a large amount of time to extract feature maps for different size of images.
\cite{wang2018automated} apply a \textit{2DFPN} network for lung nodule detection, followed by a Conditional 3-Dimensional Non-Maximum Suppression (Conditional \textit{3D-NMS}) and Attention 3D CNN (Attention 3D-CNN) for false positive reduction. However, without the spatial features within continuous CT slices, the high false positive candidates are introduced leading a great effort on the reduction process. 

\begin{figure}
    \includegraphics[width=1\textwidth]{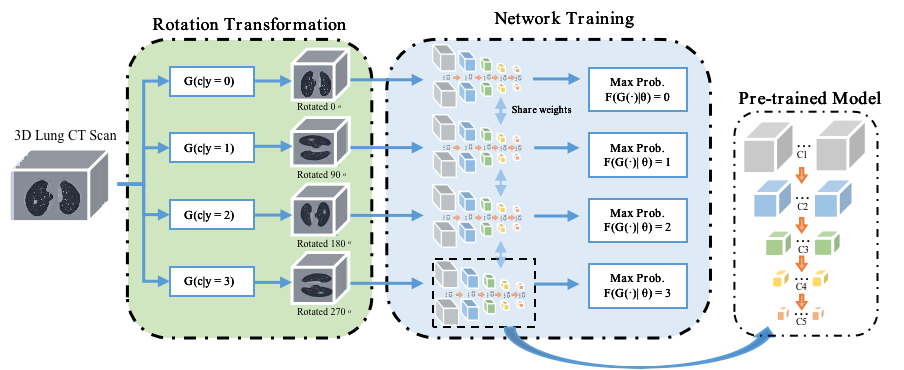}
    \caption{\textbf{The framework of the self-supervised feature learning consists of two parts.} 1) Rotation transformation on input 3D CT scans with four angles as $0^\circ,90^\circ,180^\circ,270^\circ$. 2) The rotation prediction network (\textit{ResNet-18}) is the backbone network of the proposed \textit{3DFPN} lung nodule detection framework with 2 fully connected (FC) layers to obtain the maximum prediction probabilities for each rotation angle. The trained model keeps the five ConvNets and removes two FC layers as a pre-trained model for \textit{3DFPN}.}
    \label{fig:self-supervised}
\end{figure}

In this paper, we propose a rich spatiotemporal features extraction methodology, an accurate multi-scale nodule detection network, and an efficient false-positive reduction algorithm for accurate and robust lung nodule detection.

\section{Our Method}
As shown in Figure~\ref{fig:system}, a pre-trained model is obtained from a backbone network of the nodule detection network by a self-supervised pretext task training process to extract rich spatiotemporal features to leverage the differences of CT scans captured by different brands of scanners. 
The pre-trained model is applied to the backbone network (\textit{ResNet-18}) of the lung nodule detection framework \textit{3DFPN}. The input of \textit{3DFPN} is a whole 3D volume of CT scan and the output is the 3D locations of lung nodule candidates. After the nodule candidates are detected, a 3D cube centered with the candidate is cropped and fed to a High Sensitivity and Specificity (\textit{HS$^2$}) network to further recognize whether the detected nodule candidates are real nodules or false positives. 

\subsection{Self-supervised Pre-trained Model}
To allow the proposed nodule detection framework to learn rich 3D CT features from a large-scale dataset captured by CT scanners from multiple manufacturers without additional annotations, a pre-trained model is obtained from a self-supervised pretext task training by firstly performing a four-degree rotation transformation of 3D CT scan as four individual input classes and then followed with a network for rotation prediction task to maximize the probability of rotation angle classification. Thus, rich spatiotemporal CT scans features can be learned during the process.

As shown at Figure~\ref{fig:self-supervised}, inspired by~\cite{jing2018self}, rotation transformations are applied to 3D CT scans by rotating all images in CT scans on the image plane at following four angles $\theta$ respectively {($0^\circ,90^\circ,180^\circ, 270^\circ$)}. 
For an input CT scan $c_i$ with dimensions $\{x,y,z\}$, we simply rotate it on $y$ dimension and keep the $\{x,z\}$ fixed. 
The backbone network of the proposed lung nodule detection framework (\textit{ResNet-18}) is applied to classify the rotation categories of the input CT scans with the four angles. 
Two fully connected layers are followed for probability prediction and are not used for nodule detection. 
The cross entropy loss is applied for probability prediction of $K$ rotation angles (here $k=4$) and rotation $r$ as shown in Eq. (1):
\begin{equation}
loss(c_i|\theta)= - 1/K\sum_{r=1}^Klog(F(G(c_i,r)|\theta)),
\end{equation}
where the classification network \textit{ResNet-18} is defined as $F(\cdot|\theta)$ for spatiotemporal feature learning and the rotation transformation from the input 3D CT scans to the categories of rotation angle is represented as $G(c|y)$.

\subsection{3D Feature Pyramid Network for Nodule Detection}
The recent progress in computer vision suggests feature pyramid networks (\textit{FPN}) are good at detecting objects at different scales \citep{lin2017feature}. However, the original \textit{FPNs} are designed for processing 2D images. Here, we propose a \textit{3DFPN} to detect 3D locations of lung nodules from 3D CT volumetric scans by using \textit{ResNet-18} as the backbone network which is also used in self-supervised learning to obtain the pre-trained model.
Different than~\cite{lin2017feature} which only concatenates the upper-level features in feature pyramid, we use a dense pyramid network to integrate both the low-level high-resolution features as well as high-level high-semantic features, which enriches the location details and strong semantics for nodule detection. 
Table~\ref{fpns} highlights the main differences between \textit{2DFPN} and our \textit{3DFPN}.

\begin{table}
\caption{\textbf{Comparison between \textit{2DFPN} \citep{lin2017feature} and our proposed \textit{3DFPN}.} We take 3D volume as input and the feature pyramid layers are integrated with lateral connections of all the high-level and low-level features.}\label{fpns}
\resizebox{1\textwidth}{!}{
\begin{tabular}{lccccccc}
\hline
Method  & \tabincell{c}{Input 3D\\volume} & \tabincell{c}{Lateral\\ connections} & \tabincell{c}{Integrate \\upper layer}  & \tabincell{c}{Integrate \\lower layer}   & \tabincell{c}{Upsample \\higher layer} &  \tabincell{c}{Downsample \\lower layer} \\
\hline
2DFPN \citep{lin2017feature}  & No & $\surd$ & $\surd$ &  No &$\surd$ & No\\
3DFPN (Ours)   & $\surd$ & $\surd$ &$\surd$& $\surd$  &$\surd$&$\surd$\\
\hline
\end{tabular}
}
\end{table}

\begin{figure}
    \includegraphics[width=1\textwidth]{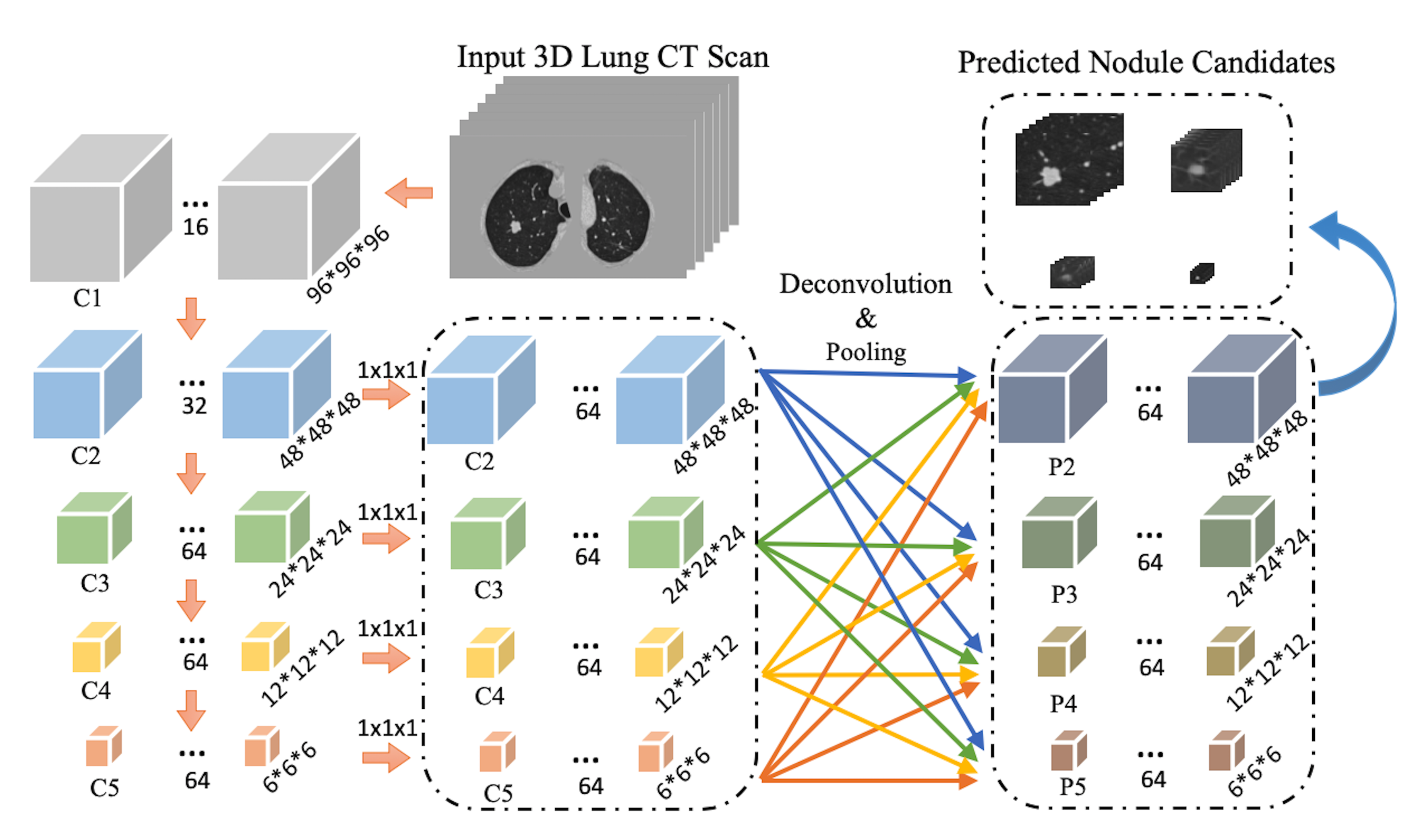}
    \caption{\textbf{Detailed illustration for the architecture of the proposed \textit{3DFPN} network.} The input 3D volume is split into $96\times96$ pixels $\times96$ slices. The size of {C1, C2, C3, C4, C5} is $96^3, 48^3, 24^3, 12^3$, and $6^3$ respectively. The following convolution layer with kernel size $1\times1\times1$ converts feature channels to 64 dimensions. 3D deconvolution and max-pooling layers are applied for integrating each of the convolution layers {C2, C3, C4, C5} to the pyramid layers {P2, P3, P4, P5}.
    } \label{fig:3DFPN}
\end{figure}

As shown in Figure~\ref{fig:3DFPN}, the bottom-up network extracts features from the convolution layers 2--5, refer as {C2, C3, C4, C5}, followed by a convolution layer with kernel size $1\times1\times1$ for converting the dimension of convolution layers to the same number. The feature pyramid network contains four layers {P2, P3, P4, P5}, which integrates the low-level features by a max-pooling layer and the downsampled high-level features by a 3D deconvolution layer. \textit{3DFPN} predicts a confidence score for each nodule candidate and the corresponding location with four parameters as $[x,y,z,d]$, where $[x,y]$ are the spatial coordinates at each CT slice, $z$ is the CT slice index number, $d$ as nodule diameter.

\subsection{HS$^2$ Network for False Positive Reduction}

\begin{figure}
    \includegraphics[width=1\textwidth]{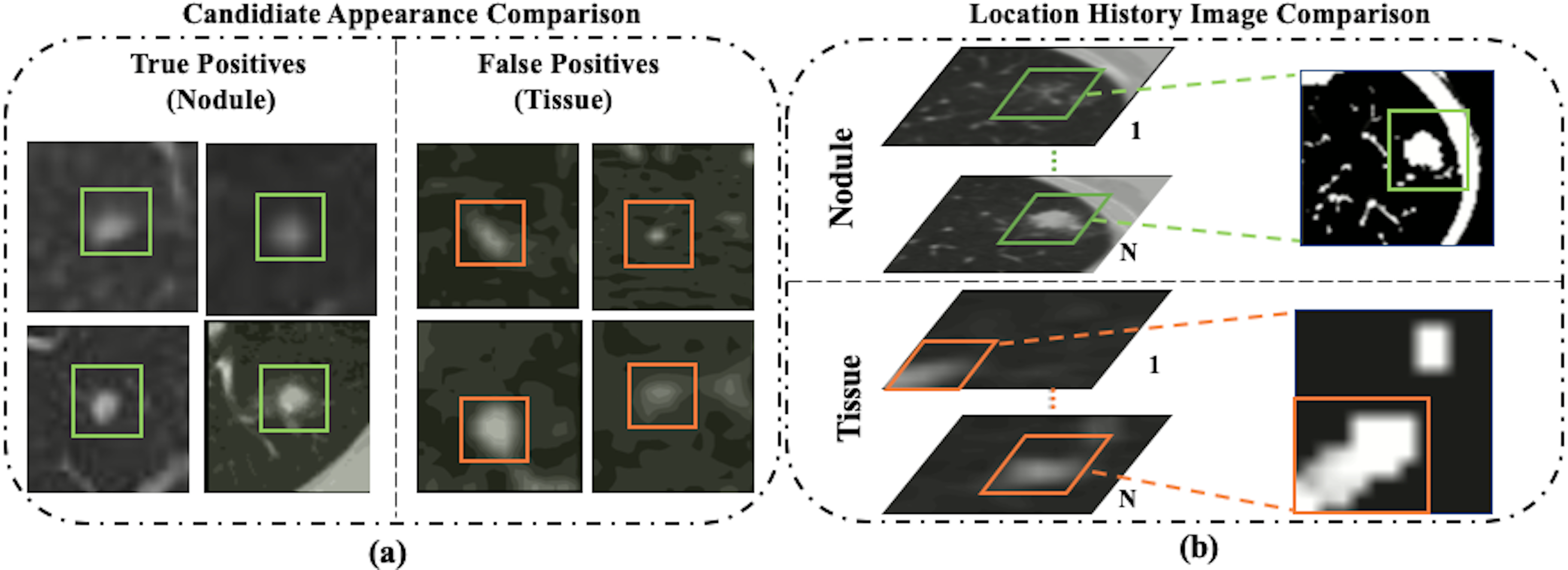}
    \centering
    \caption{\textbf{The proposed Location History Images (LHI) to distinguish tissues and nodules from the predicted nodule candidates.} (a) The similar appearance of true nodules (green boxes) and false detected tissues (orange boxes). (b) The orientations of the location variances for nodules and tissues are presented differently in LHIs. True nodules generally have a circular region which indicates the spatial changes with either a brighter center (when nodule sizes in following CT slices are smaller) or a darker center (when nodule sizes in following CT slices are bigger). On the other hand, the location variance for false detected tissues usually tends to change in certain directions such as a gradually changed trajectory line. 
    } \label{fig:candiates}
\end{figure}

Due to the low resolution and the noise of CT images, as shown in Figure~\ref{fig:candiates}(a), the appearances of some tissues (orange boxes) are similar to real nodules (green boxes) which are very likely to be detected as nodule candidates by \textit{3DFPN} and leads to a large number of false positives. 
As shown in Table~\ref{nodulestats}, we further analyze 300 false positives predicted by the \textit{3DFPN} and observe that 241 False Positives (FPs) are caused by the high appearance similarity of tissues ($80.3\%$), 33 of them are caused by inaccurate size detection ($11\%$), and 26 FPs are due to inaccurate location detection ($8.7\%$).

\begin{figure}
    \includegraphics[width=.9\textwidth]{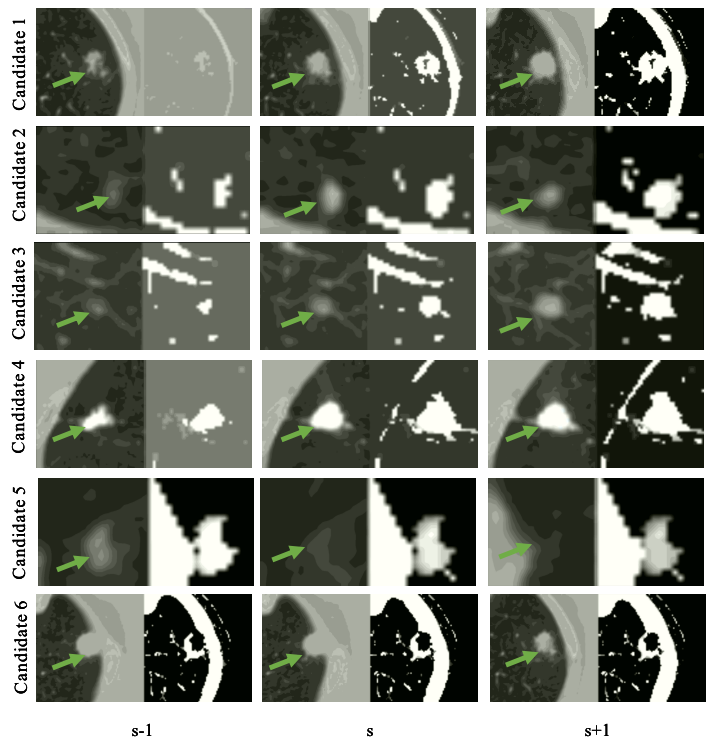}
    \centering
    \caption{\textbf{Examples of the detected true nodule candidates (the left image of each column) and their corresponding LHIs (the right image of each column) calculated between ($s-2$, $s-1$), ($s-1$, $s$), and ($s$, $s+1$) slices shown in the $s-1$, $s$, $s+1$ columns.} The green arrows mark the position of candidates. As shown in the figure, the true nodules have a circular region on LHI images as the location of the nodule approach to the center or the edge of nodule volume. In addition, the center location of the nodule candidates barely changes in the continuous slices.
    } \label{fig:LHINodule}
    
\end{figure}

\begin{figure}
    \includegraphics[width=.8\textwidth]{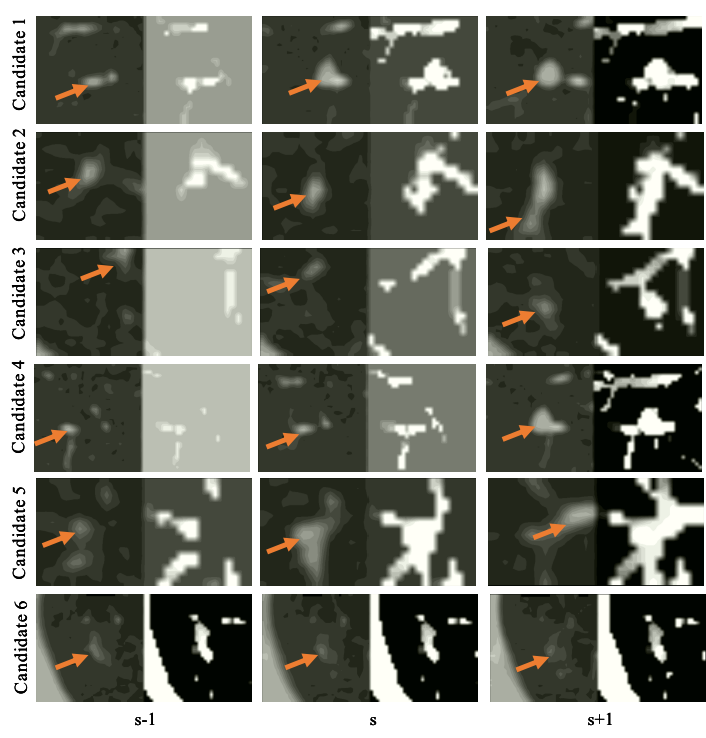}
    \centering
    \caption{\textbf{Examples of false detected tissue candidates (the left image of each column) and their corresponding LHIs (the right image of each column) calculated between three continuous slices ($s-2$, $s-1$), ($s-1$, $s$), and ($s$, $s+1$) shown in the $s-1$, $s$, $s+1$ columns.} The orange arrows mark the position of false detected tissue candidates. LHIs of tissues are shown to have clear differences with true nodules. Compared with the LHIs of the true nodules in Figure~\ref{fig:LHINodule}, the wide variation of tissue location follows certain patterns, illustrated as intensity variances along the trajectory lines in the LHIs. 
    } \label{fig:LHITissue}
\end{figure}

Since the majority of the false positives are caused by tissue regions with similar appearance, it is crucial to design a method to distinguish them from nodules for false positive reduction. By treating each CT scan as a video, we discover that the orientation of the location changes could be tracked in certain patterns for tissues among continues slices, while the variance of true nodules tends to expand outside the contour or diminishing to the center at continuous CT slices as shown in  Figure~\ref{fig:candiates}(b), Figure~\ref{fig:LHINodule} and Figure~\ref{fig:LHITissue}.

\begin{table}
\caption{\textbf{Statistic Analysis for False Positive Nodule Candidates.}}
\resizebox{.8\textwidth}{!}{
\begin{tabular}{lccc}
\hline
& Tissue & Inaccurate Size & Inaccurate Location \\
\hline
Percentage& 80.3\% & 11\% & 8.7\%  \\
\hline
\end{tabular}}
\label{nodulestats}
\end{table}


Inspired by Motion History Image (MHI)~\citep{davis2001hierarchical,yang2012recognizing}, we define the Location History Image (LHI) as $f$. By given any pixel location $(x, y)$ on a CT slice $s$, $f(x, y, s)$ represents the intensity value of LHI within $(1, \tau)$ slice. The LHI is fed to an \textit{HS$^2$} (high sensitivity and specificity) network which is a feed-forward neural network with 2 convolution layers followed by 3 fully connected layers. The outputs of the \textit{HS$^2$} network are refined predicted labels of true nodules and tissues. Sensitivity is defined as a ratio of true positives over the total number of true positives and false negatives. Specificity is the ratio of true negatives over the total number of true negatives and false positives.

The intensity of LHI is calculated as in Eq. (2):
\begin{equation}
\centering
f(x,y,s)=\left\{
\begin{array}{lcl}
\tau     &      & {if~~\psi(x, y, s) = 1} \\
max(0, f(x, y, s-1)-1)    &      & {otherwise}
\end{array} \right. 
\end{equation}
where the update function $\psi(x, y, s)$ given by the spatial differentiation of the pixel intensity of two continuous CT slices. 
The algorithm has the following steps. 1) If $| I(x, y, s) - I(x, y, s-1)|$ is larger than a threshold, $\psi(x, y, s) = 1$, otherwise, $\psi(x, y, s) = 0$. 2) For the current slice, if $\psi(x,y, s) = 1$, $f = \tau$. Otherwise, if $f(x, y, s)$ is not zero, it is attenuated with a gradient of 1. If $f(x, y, s)$ equals zero, then remains as zero. 3) Repeat steps 1) and 2) until all the slices are processed.
Therefore, the location variance among continues CT slices and their change patterns can be effectively represented by our proposed LHIs. 

\section{Experimental Results and Discussions}
\subsection{Datasets}

In this paper, we employ five publicly available datasets for training, testing, and performance evaluation across different datasets: NLST, LUNA16, SPIE-AAPM, LungTime, and HMS Lung Cancer datasets. Table~\ref{table:datasets} summarizes the details of these datasets.

\begin{table}
\footnotesize
\caption{\textbf{Detailed Information of the Datasets.}}
\vspace{0.2pt}
\label{table:datasets}
\resizebox{.9\textwidth}{!}{
\begin{tabular}{lccccc}
\hline
Dataset &  Manufacturer  & CT number & Nodule number \\
\hline
NLST& \tabincell{c}{GE Healthcare\\Philips Healthcare\\Siemens Healthcare\\Toshiba Healthcare} & 13,762 & - \\

LUNA16& GE Healthcare &  888 &  1186 \\

SPIE-AAPM & Philips Healthcare & 70 & 70  \\

LungTIME &  Siemens Healthcare & 157 & 394  \\

HMS Lung Cancer & GE Healthcare & 229 & 254 \\
\hline
\end{tabular}} 
\end{table}

\textit{\textbf{NLST} Dataset:} The national lung screening trial (NLST)~\citep{national2011national} is a public dataset aimed to determine whether low-dose spiral CT screening for lung cancer can reduce lung cancer mortality in high-risk populations compared with chest screening. The data includes participant characteristics, screening test results, diagnostic procedures, lung cancer and mortality with more than 75,000 CT scans captured by four different manufacturers of CT scanners (i.e. GE, Philips, Siemens, and Toshiba.) Since this dataset is large and there are no annotations for nodule locations, the NLST dataset is used for rotation transformation based self-supervised feature learning. In a total of 13,762 CT scans are applied, where 12,000 CT scans are for training in this paper and 1,762 CT scans are for testing.

\textit{\textbf{LUNA16} Dataset:} LUNA16 challenge dataset~\citep{Aaa2017Validation} consists of $1,186$ nodules in the size between 3-30 mm from $888$ CT scans and agreed by at least 3 out of 4 radiologists. It is the most widely used dataset for performance evaluation and is divided into 10 subsets. It is used for fine-tuning the \textit{3DFPN} based on the self-supervised pre-trained model from NLST dataset and performance evaluation. To conduct a fair comparison with other methods for lung nodule detection, we follow the same protocol of cross-validation by applying 9 subsets as training and the remaining subset as testing and then report the average performance.

\textit{\textbf{SPIE-AAPM} Dataset:} The \textit{SPIE-AAPM} dataset is collected for a 'Grand Challenge' of the diagnostic classification of malignant and benign lung nodule by the international society for optics and photonics (SPIE) with the support of American Association of Physicists in Medicine (AAPM) and the National Cancer Institute (NCI)~\citep{armato2015spie}. It contains 70 CT scans from 70 patients, with the annotation of nodule location as well as the nodule diagnosis categories of benign or malignant. It is employed in our paper for cross-dataset testing.

\textit{\textbf{Lung TIME} Dataset:} The Lung Test Images from Motol Environment (Lung TIME) is publicly available and contains 157 CT scans with 394 nodules~\citep{dolejsi2009lung}. The nodules are in the range of 2-10 mm in diameter. The CT scans annotations of the nodule location are provided. It is employed in our paper for cross-dataset testing.

\textit{\textbf{HMS Lung Cancer} Dataset:} HMS Lung Cancer dataset~\citep{mak2019use} contains the CT scans and lung tumor sections generated by clinical care professionals used in competition with 461 patients. HMS contains a total number of 229 CT scans and 254 nodules with the nodule location annotation. It is employed in our paper for cross-dataset testing.

\subsection{Data Preprocessing}  A preprocessing procedure is required to original CT scans for accurate nodule detection. First, the masks of the lung regions are extracted by lung region segmentation. By using the traditional method, the 2D single slice is processed first with a Gaussian filter to remove the fat, water and kidney background and followed by a 3D connection volume extraction to remove unrelated areas~\citep{liao2019evaluate}. However, it takes $9$ to $22$ seconds to obtain the mask for each CT scan. In order to accelerate the processing speed for large datasets, we employ the LGAN method ~\citep{tan2019lgan} and train the network on $10,000$ CT slices for lung mask extraction to speed up the process in an average of $5$ seconds per scan. In addition, CT scans with an effective value of Hounsfield unit between $[-1200, 600]$ are transformed into the gray value of $[0, 255]$ by a linear mapping. The spacing (mm/pixel) of CT scans between different patients and machines are various and the re-sampling is applied to unify the spacing to 1 mm.

\subsection{Experimental Settings} 
\subsubsection{Self-supervised Pre-trained Model}
The self-supervised pretext task takes the 3D CT images of four rotation angles at $0,90,180,270$ degrees as input to learn the spatiotemporal features of CT scans. Followed by~\cite{jing2018self}, we adopt the flip and transpose for the rotation transformation. The backbone network of \textit{3DFPN} (\textit{ResNet-18}) is adopted to predict the geometric transformation. The five ConvNets followed with two fully connected layers are trained to maximize the probability of input rotation angle by a four-way classification task. The pre-trained model without the 2 fully connected layers is then employed as the backbone network \textit{3DFPN} of for lung nodule detection.

During the training, the learning rate is set to 0.1, decreasing by $1/2$ after 70 and 85 epochs, with the weight decay of $5e^{-4}$. The total training includes 100 epochs and the batch size is set to 16.

\subsubsection{3DFPN}
The framework takes the whole CT scan as input, while volume at $96\times96\times96$ pixels is selected by a sliding window method as the input of the \textit{3DFPN} network. This size is selected based on experiments to ensure it is big enough to contain the whole nodule even when it is with the largest size (about 30 mm). The anchor sizes employed in our \textit{3DFPN} to obtain the candidate regions from feature maps are [3$^3$, 5$^3$, 10$^3$, 15$^3$, 20$^3$, 25$^3$, 30$^3$] pixels. For all the anchors, the corresponding regions obtained from all the 3D feature map levels are gathered to predict the position of nodules.

In the training phase, the regions with an Intersection-over-Union (IoU) threshold to the ground-truth regions less than $0.02$ are referred to negative samples and greater than $0.4$ are positive samples. The regions with IoU value in between are ignored to avoid the similarity of the positive and negative samples. A classification layer is used to predict a confidence score for the candidate class and a region regression layer is applied to learn the offset between the position of region proposals and the ground-truth. We adopt Smooth $L1$ loss \citep{girshick2015fast} and binary cross entropy loss (BCE-loss) for location regression and classification score respectively. 
In the testing, for each region proposal, a confidence score is calculated by the classification layer. The proposals with a probability larger than 0.1 are chosen as nodule candidates. Non-maximum suppression is further applied to eliminate the multiple predicted candidates for one nodule.

\subsubsection{HS$^2$ network} 
The two convolution layers of the \textit{HS$^2$} network are set to $(1, 30)$, $(30, 50)$ dimensions and followed by three fully connected layers with the channel sizes of $(2048, 1024, 512)$. The cross entropy loss is applied for classification during the training. Image patches aligned with each predicted nodule candidate region but with twice size (in both x and y directions) are selected from $11$ continuous CT slices ($5$ slices before and after the current slice of nodule candidate respectively). The LHIs of these patches are computed and resized to $48\times48$ pixels as the input of the \textit{HS$^2$} network. 

In the training, the learning rate starts from $0.01$ and decreases to $1/10$ for every $500$ epochs. Total of $2,000$ epochs is conducted for the framework. The average prediction time for a whole CT scan is about $0.53$ min/scan on a server with one GeForce GTX 1080 GPU using Pytorch 2.7. 

\subsection{Evaluation Metrics}
In order to conduct a fair comparison with other methods on LUNA16 dataset, we follow the same process to conduct cross validations by using 9 subsets for training and the remaining 1 subset for testing, then obtain the final results by averaging the 10 experiments. Data augment is applied by flipping and resizing the CT scans. 
Same as other methods, the Free-Response Receiver Operating Characteristic (FROC) analysis \citep{Setio2016Pulmonary} and Competition Performance Metric (CPM) are employed to measure the performance. Follow the LUNA16 challenge evaluation method, the FROC curve plots detection sensitivity and the corresponding false positives where the points on the curve are obtained by the true positive rate (true positives over the sum of true positives and false negatives) while the false positive rate at $1/8, 1/4, 1/2, 1, 2, 4, 8$ per scan respectively. The CPM score is calculated by the average of sensitivity for all the levels of false positives per scan.

\subsection{Experimental Results and Analysis}
\subsubsection{Comparison with Other Methods} 
\begin{figure}
\centering
\includegraphics[width=.7\textwidth]{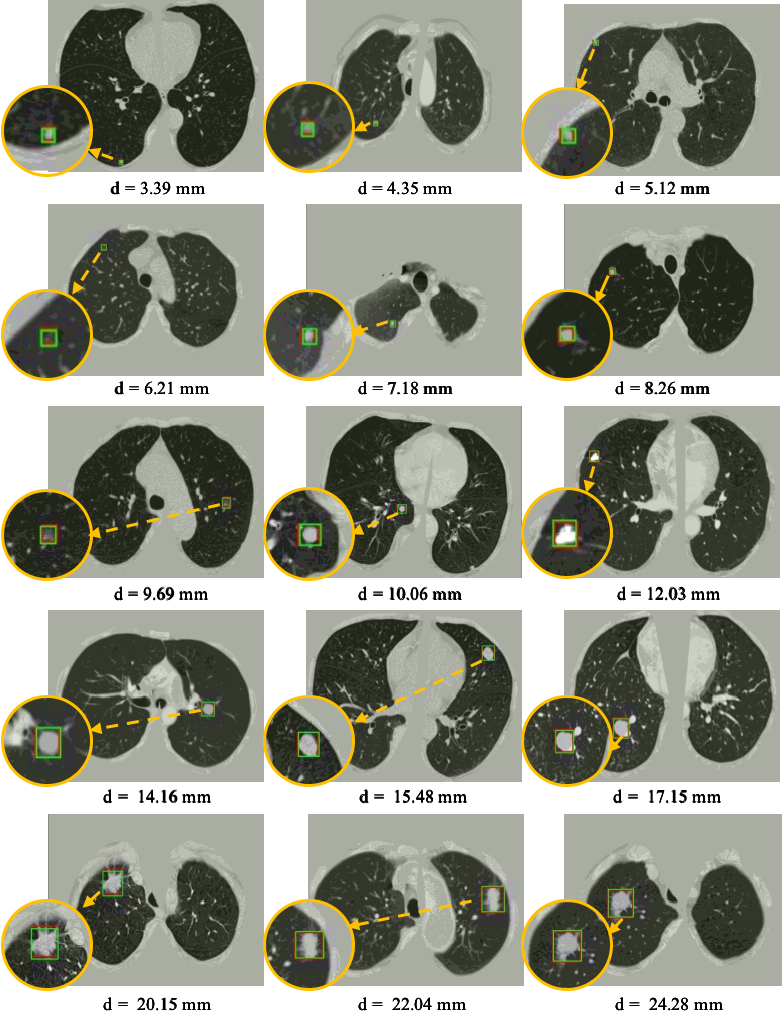}
\caption{\textbf{Visualization of some detected true nodules with different sizes from 3mm to 25mm in diameter $d$ by our proposed \textit{3DFPN-HS$^2$} framework.} For a better visualization, the detected nodule regions are zoomed in as shown in the orange circles. The green box indicates the predicted region and the red box represents the ground-truth. Some of the red boxes are not clearly observed because they are perfectly overlapped with the green boxes. The results demonstrate that our \textit{3DFPN-HS$^2$} framework is capable to accurately detect lung nodules of different sizes from CT scans.} 
\label{fig:MoreResult}
\end{figure}

Table~\ref{tab1e:overall} shows the FROC evaluation results of our proposed method compared with state-of-the-art methods on LUNA16 at $1/8,1/4,1/2,1,2,4,8$ false positive levels. The highlighted numbers in the table indicate the best performance within each column. Since the pre-trained model is not used in most state-of-the-art methods, for a fair comparison, all the methods are tested on LUNA16 dataset followed the same FROC evaluation without the pre-trained model. As shown in the table, our framework outperforms $5.5\%$ average sensitivity than the best result of other methods. In addition, the proposed framework achieves the best performance at every FP level. As previously mentioned, the CAD system is not only required a high sensitivity, but also high specificity. Table~\ref{tab1e:overall} demonstrates that the false positives are greatly reduced by the proposed \textit{HS$^2$} network. \textit{3DFPN-HS$^2$} obtains a highest $97.14\%$ sensitivity at $2$ FPs per scan. In addition, for the FP of $1/8$, $1/4$, and $1/2$ per scan, the proposed framework still remains a high sensitivity above $90\%$. The experimental results show that \textit{3DFPN-HS$^2$} reaches a state-of-the-art performance for high sensitivity and specificity lung nodule detection. The illustration of nodule detection results is shown in Figure~\ref{fig:MoreResult}.

\begin{table}
\caption{\textbf{FROC Performance comparison with the state-of-the-art methods on LUNA16 dataset}: sensitivity (recall) and the corresponding false positives at $1/8, 1/4, 1/2, 1, 2, 4, 8$ per scan. Our \textit{3DFPN-HS$^2$} method achieves the best performance (with $>90\%$ sensitivity) at all false positive levels and significantly outperforms others especially at the low false positive levels ($1/8$ and $1/4$). \textit{3DFPN-HS$^2\star$} shows the result with the pre-trained self-supervised feature model from NLST dataset.}
\resizebox{1\textwidth}{!}{
\begin{tabular}{lcccccccc}
\hline
\centering {\bfseries Methods} &  {\bfseries1/8 } & {\bfseries1/4 }& {\bfseries1/2 }&{\bfseries1 }& {\bfseries2} & {\bfseries4} & {\bfseries8} & {\bfseries CPM score}\\
\hline
\cite{gupta2018automatic}& 0.531&0.629&0.790&0.835&0.843&0.848&0.856&0.763\\

\cite{dou2017multilevel}&0.677& 0.737 &0.815 &0.848 &0.879 &0.907 &0.922& 0.827\\

\cite{dou2017automated} & 0.659 & 0.745 & 0.819 & 0.865 & 0.906 &0.933&	0.946&	0.839\\

\cite{zhu2018deeplung}& 0.692 & 0.769 & 0.824 & 0.865 & 0.893 & 0.917& 0.933 & 0.842\\

\cite{wang2018automated}& 0.676 & 0.776 & 0.879 & 0.949 & 0.958 & 0.958& 0.958 & 0.878\\

\cite{Ding2017Accurate}& 0.748 &0.853 &0.887& 0.922& 0.938& 0.944 &0.946& 0.891\\
 
\cite{khosravan2018s4nd} & 0.709& 	0.836&	0.921& 0.953& 0.953& 0.953& 0.953& 0.897\\
 \hline
 {\bfseries 3DFPN (Ours)} & {\bfseries 0.848} & {\bfseries 0.876} & {\bfseries 0.905} & {\bfseries 0.933}& {\bfseries0.943}&{\bfseries0.957}&{\bfseries0.970}& {\bfseries0.919}\\

 {\bfseries 3DFPN-HS$^2$ (Ours)} & {\bfseries 0.904} & {\bfseries 0.914} &  {\bfseries0.933} & {\bfseries0.957} &  {\bfseries0.971} &{\bfseries0.971}&{\bfseries0.971}& {\bfseries0.952}\\

 {\bfseries 3DFPN-HS$^2\star$ (Ours)} &{\bfseries 0.906} & {\bfseries 0.923} &  {\bfseries0.948} & {\bfseries0.981} &  {\bfseries0.981} &{\bfseries0.981}&{\bfseries0.981}& {\bfseries0.957}\\
 \hline
\end{tabular}}
\label{tab1e:overall}
\end{table}

\subsubsection{Robustness with the Self-supervised Pre-trained Model}
As shown in Table~\ref{table:pretrain}, two sets of experiments are conducted to evaluate the robustness of the proposed framework across different datasets without and with the self-supervised based pre-trained model respectively.  One set uses the \textit{3DFPN-HS$^2$} model trained only on LUNA16 training data without using the self-supervised based pre-trained model from NLST dataset. Another set applies the pre-trained model and then fine-tuned on LUNA16 dataset which the results are highlighted in the table.
All the results are tested on LUNA16 testing data, SPIE-AAPM, LungTime, and HMS Lung Cancer datasets. 
The results with the pre-trained model on LUNA16 test set show slight improvements than the results without using the pre-trained model at all the false positive levels . 
For the testing on SPIE-AAPM, LungTime, and HMS Lung Cancer datasets which are never used in training and fine-tuning, the performance significantly decreases compared to the performance on LUNA16 especially at $1/8$ false positive per scan without using the pre-trained model. This is because the training set of LUNA16 is relatively small to obtain a robust \textit{3DFPN-HS$^2$} model. 
Compared with the model trained only on the LUNA16 data set, the framework with pre-trained model showed significant improvement in all false positive levels across all these datasets. In particular, on $1/8$ false positive per scan, the sensitivity is boosted up to $7.4\%$ on SPIE-AAPM, $13.5\%$ on LungTIME, and $8.9\%$ on HMS respectively. For the 8 false positives per scan, the accuracy is equivalent to that of LUNA16. 
The significant improvement in performance demonstrates the robustness of applying the self-supervised pre-trained model across different datasets without requiring additional annotations. 

\begin{table}
\caption{\textbf{FROC Performance comparison with and without using the self-supervised learning-based pre-trained modelz}: sensitivity (recall) and the corresponding false positives at $1/8, 1/4, 1/2, 1, 2, 4, 8$ per scan. The sensitivities at all false positive levels with the pre-trained model for the new datasets significantly outperform the performance without the pre-trained model. Note: The pre-trained model of CT features is obtained by predicting 4 rotations on NLST dataset. The \textit{3DFPN-HS$^2$} for lung nodule detection is only finetuned (with the pre-trained model)/trained (without the pre-trained model) on LUNA16 dataset.}
\vspace{0.2cm}
\resizebox{1\textwidth}{!}{
\begin{tabular}{lccccccccc}
\hline
{\bfseries Datasets} & {\bfseries Pretrained model} & {\bfseries1/8 } & {\bfseries1/4 }& {\bfseries1/2 }&{\bfseries1 }& {\bfseries2} & {\bfseries4} & {\bfseries8} & {\bfseries CPM score}\\
\hline
\multirow{2}*{LUNA16} & no & 0.904 & 0.914 & 0.933 & 0.957 &  0.971 &0.971&0.971& 0.952\\

~ & $\surd$ & {\bfseries 0.906} & {\bfseries 0.923} &  {\bfseries0.948} & {\bfseries0.981} &  {\bfseries0.981} &{\bfseries0.981}&{\bfseries0.981}& {\bfseries0.957}\\
\hline
 \multirow{2}*{SPIE-AAPM} & no & 0.734& 0.771 &0.809& 0.863& 0.895& 0.914& 0.923 & 0.844\\

 ~ &$\surd$ & {\bfseries 0.808} & {\bfseries 0.852} & {\bfseries 0.877} & {\bfseries 0.914}& {\bfseries0.928}&{\bfseries0.947}&{\bfseries0.964}& {\bfseries0.899}\\
 \hline
  \multirow{2}*{LungTIME} &no & 0.662 & 0.746 & 0.815 & 0.864 & 0.902 &0.918 & 0.932& 0.834\\

 ~ &$\surd$ & {\bfseries 0.797} & {\bfseries 0.826} & {\bfseries 0.862} & {\bfseries 0.891}& {\bfseries0.927}&{\bfseries0.948}&{\bfseries0.969}& {\bfseries0.889}\\
 \hline
  \multirow{2}*{HMS Lung Cancer} & no & 0.657 & 0.703 & 0.746 & 0.805 & 0.846 & 0.912& 0.927 & 0.799\\

  ~ & $\surd$ & {\bfseries 0.746} & {\bfseries 0.816} & {\bfseries 0.842} & {\bfseries 0.889}& {\bfseries0.909}&{\bfseries0.937}&{\bfseries0.956}& {\bfseries0.871}\\
\hline
\end{tabular}}
\label{table:pretrain}
\end{table}

\begin{figure}
    \centering
    \includegraphics[width=.9\textwidth]{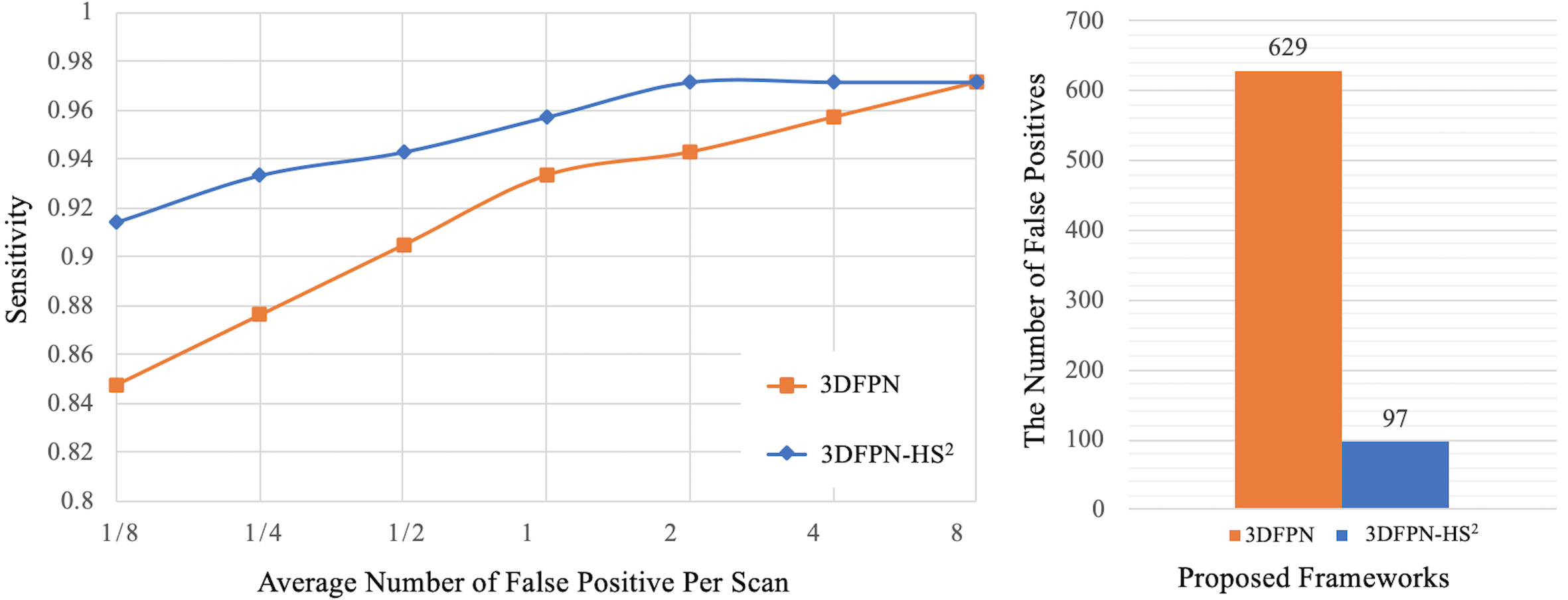}
    \caption{\textbf{Comparison between the proposed \textit{3DFPN} and \textit{3DFPN-HS$^2$}.} Left: Comparison between the proposed \textit{3DFPN} and \textit{3DFPN-HS$^2$} (with High Sensitivity and Specificity network for false positive reduction) on LUNA16 dataset without using the self-supervised pre-trained model. \textit{3DFPN-HS$^2$} greatly improves the performance of the \textit{3DFPN} at all the FP levels. Right: The number of false positives is reduced from $629$ to $97$ for a total of $88$ CT scans with the confidence score above $0$ after the \textit{HS$^2$} network is applied.
    } \label{fig:Result1}
\end{figure}

\subsubsection{Effectiveness of HS$^2$ Network for FP Reduction} Two experiments are conducted to demonstrate the advantages of \textit{HS$^2$} network on LUNA16 dataset. As shown in Figure~\ref{fig:Result1} (left), compared with \textit{3DFPN} without the \textit{HS$^2$} network, the result of \textit{3DFPN-HS$^2$} with the false positive reduction is increased more than 5\% at 1/8 FP level. In addition, the numbers of FPs with (blue bar) and without (orange bar) \textit{HS$^2$} network for all the predicted nodule candidates from a total of 88 CT scans (subset 9) are further compared in  Figure~\ref{fig:Result1} (right). By applying \textit{HS$^2$}, the \textit{3DFPN-HS$^2$} is able to distinguish the false detected tissues from true nodules, therefore significantly reduces FPs by 84.5\%. It is worth noting that our proposed \textit{3DFPN} without \textit{HS$^2$} network still achieves 97\% at 8 FPs per scan and 91.9\% CPM, which surpasses other state-of-the-art methods (see Table~\ref{tab1e:overall}.)

\section{Conclusion}
In this paper, we have proposed an effective and robust \textit{3DFPN-HS$^2$} framework for lung nodule detection with a self-supervised feature learning schema. By employing a 3D feature pyramid network with local and global feature enrichment, different sizes of lung nodules can be detected. By introducing the \textit{HS$^2$} network and treating each CT scan as a video, false positives are significantly reduced based on the different patterns of location variance for nodules and tissues in continuous CT slices. By applying a self-supervised feature learning schema, spatiotemporal features of CT scans can be effectively learned from large-scale CT scans without using additional labels. The learned features can significantly improve the robustness of the proposed framework across different clinic datasets.  The proposed framework strikingly outperforms state-of-the-art methods, has achieved both high sensitivity and specificity and is robust to the data from multiple CT scanner manufacturers which has great potential in routine clinical practice. 

\section*{Acknowledgements}
This material is based upon work supported by the National Science Foundation under award number IIS-1400802 and Memorial Sloan Kettering Cancer Center Support Grant/Core Grant P30 CA008748. Oguz Akin, MD serves as a scientific advisor for Ezra AI, Inc., which is unrelated to the research being reported.


\bibliography{mybibfile}

\begin{thebibliography}{47}
\expandafter\ifx\csname natexlab\endcsname\relax\def\natexlab#1{#1}\fi
\providecommand{\url}[1]{\texttt{#1}}
\providecommand{\href}[2]{#2}
\providecommand{\path}[1]{#1}
\providecommand{\DOIprefix}{doi:}
\providecommand{\ArXivprefix}{arXiv:}
\providecommand{\URLprefix}{URL: }
\providecommand{\Pubmedprefix}{pmid:}
\providecommand{\doi}[1]{\href{http://dx.doi.org/#1}{\path{#1}}}
\providecommand{\Pubmed}[1]{\href{pmid:#1}{\path{#1}}}
\providecommand{\bibinfo}[2]{#2}
\ifx\xfnm\relax \def\xfnm[#1]{\unskip,\space#1}\fi
\bibitem[{Aaa et~al.(2017)Aaa, Traverso, De, Msn, Cvd, Cerello, Chen, Dou,
  Fantacci and Geurts}]{Aaa2017Validation}
\bibinfo{author}{Aaa, S.}, \bibinfo{author}{Traverso, A.}, \bibinfo{author}{De,
  B.T.}, \bibinfo{author}{Msn, B.}, \bibinfo{author}{Cvd, B.},
  \bibinfo{author}{Cerello, P.}, \bibinfo{author}{Chen, H.},
  \bibinfo{author}{Dou, Q.}, \bibinfo{author}{Fantacci, M.E.},
  \bibinfo{author}{Geurts, B.}, \bibinfo{year}{2017}.
\newblock \bibinfo{title}{Validation, comparison, and combination of algorithms
  for automatic detection of pulmonary nodules in computed tomography images:
  The luna16 challenge.}
\newblock \bibinfo{journal}{Medical Image Analysis} \bibinfo{volume}{42},
  \bibinfo{pages}{1--13}.
\newblock \bibinfo{note}{[dataset]}.
\bibitem[{Armato~III et~al.(2015)Armato~III, Hadjiiski, Tourassi, Drukker,
  Giger, Li, Redmond, Farahani, Kirby and Clarke}]{armato2015spie}
\bibinfo{author}{Armato~III, S.G.}, \bibinfo{author}{Hadjiiski, L.},
  \bibinfo{author}{Tourassi, G.}, \bibinfo{author}{Drukker, K.},
  \bibinfo{author}{Giger, M.}, \bibinfo{author}{Li, F.},
  \bibinfo{author}{Redmond, G.}, \bibinfo{author}{Farahani, K.},
  \bibinfo{author}{Kirby, J.}, \bibinfo{author}{Clarke, L.},
  \bibinfo{year}{2015}.
\newblock \bibinfo{title}{Spie-aapm-nci lung nodule classification challenge
  dataset}.
\newblock \bibinfo{journal}{Cancer Imaging Arch} \bibinfo{volume}{10},
  \bibinfo{pages}{K9}.
\newblock \bibinfo{note}{[dataset]}.
\bibitem[{Ciompi et~al.(2015)Ciompi, de~Hoop, van Riel, Chung, Scholten,
  Oudkerk, de~Jong, Prokop and van Ginneken}]{ciompi2015automatic}
\bibinfo{author}{Ciompi, F.}, \bibinfo{author}{de~Hoop, B.},
  \bibinfo{author}{van Riel, S.J.}, \bibinfo{author}{Chung, K.},
  \bibinfo{author}{Scholten, E.T.}, \bibinfo{author}{Oudkerk, M.},
  \bibinfo{author}{de~Jong, P.A.}, \bibinfo{author}{Prokop, M.},
  \bibinfo{author}{van Ginneken, B.}, \bibinfo{year}{2015}.
\newblock \bibinfo{title}{Automatic classification of pulmonary peri-fissural
  nodules in computed tomography using an ensemble of 2d views and a
  convolutional neural network out-of-the-box}.
\newblock \bibinfo{journal}{Medical image analysis} \bibinfo{volume}{26},
  \bibinfo{pages}{195--202}.
\bibitem[{Davis(2001)}]{davis2001hierarchical}
\bibinfo{author}{Davis, J.W.}, \bibinfo{year}{2001}.
\newblock \bibinfo{title}{Hierarchical motion history images for recognizing
  human motion}, in: \bibinfo{booktitle}{Proceedings IEEE Workshop on Detection
  and Recognition of Events in Video}, pp. \bibinfo{pages}{39--46}.
\bibitem[{Ding et~al.(2017)Ding, Li, Hu and Wang}]{Ding2017Accurate}
\bibinfo{author}{Ding, J.}, \bibinfo{author}{Li, A.}, \bibinfo{author}{Hu, Z.},
  \bibinfo{author}{Wang, L.}, \bibinfo{year}{2017}.
\newblock \bibinfo{title}{Accurate pulmonary nodule detection in computed
  tomography images using deep convolutional neural networks}, in:
  \bibinfo{booktitle}{International Conference on Medical Image Computing and
  Computer-Assisted Intervention}, \bibinfo{organization}{Springer}. pp.
  \bibinfo{pages}{559--567}.
\bibitem[{Doersch et~al.(2015)Doersch, Gupta and
  Efros}]{doersch2015unsupervised}
\bibinfo{author}{Doersch, C.}, \bibinfo{author}{Gupta, A.},
  \bibinfo{author}{Efros, A.A.}, \bibinfo{year}{2015}.
\newblock \bibinfo{title}{Unsupervised visual representation learning by
  context prediction}, in: \bibinfo{booktitle}{Proceedings of the IEEE
  International Conference on Computer Vision}, pp.
  \bibinfo{pages}{1422--1430}.
\bibitem[{Doersch and Zisserman(2017)}]{doersch2017multi}
\bibinfo{author}{Doersch, C.}, \bibinfo{author}{Zisserman, A.},
  \bibinfo{year}{2017}.
\newblock \bibinfo{title}{Multi-task self-supervised visual learning}, in:
  \bibinfo{booktitle}{Proceedings of the IEEE International Conference on
  Computer Vision}, pp. \bibinfo{pages}{2051--2060}.
\bibitem[{Dolejsi et~al.(2009)Dolejsi, Kybic, Polovincak and
  Tuma}]{dolejsi2009lung}
\bibinfo{author}{Dolejsi, M.}, \bibinfo{author}{Kybic, J.},
  \bibinfo{author}{Polovincak, M.}, \bibinfo{author}{Tuma, S.},
  \bibinfo{year}{2009}.
\newblock \bibinfo{title}{The lung time: Annotated lung nodule dataset and
  nodule detection framework}, in: \bibinfo{booktitle}{Medical Imaging 2009:
  Computer-Aided Diagnosis}, \bibinfo{organization}{International Society for
  Optics and Photonics}. p. \bibinfo{pages}{72601U}.
\newblock \bibinfo{note}{[dataset]}.
\bibitem[{Dou et~al.(2017a)Dou, Chen, Jin, Lin, Qin and
  Heng}]{dou2017automated}
\bibinfo{author}{Dou, Q.}, \bibinfo{author}{Chen, H.}, \bibinfo{author}{Jin,
  Y.}, \bibinfo{author}{Lin, H.}, \bibinfo{author}{Qin, J.},
  \bibinfo{author}{Heng, P.A.}, \bibinfo{year}{2017}a.
\newblock \bibinfo{title}{Automated pulmonary nodule detection via 3d convnets
  with online sample filtering and hybrid-loss residual learning}, in:
  \bibinfo{booktitle}{International Conference on Medical Image Computing and
  Computer-Assisted Intervention}, pp. \bibinfo{pages}{630--638}.
\bibitem[{Dou et~al.(2017b)Dou, Chen, Yu, Qin and Heng}]{dou2017multilevel}
\bibinfo{author}{Dou, Q.}, \bibinfo{author}{Chen, H.}, \bibinfo{author}{Yu,
  L.}, \bibinfo{author}{Qin, J.}, \bibinfo{author}{Heng, P.A.},
  \bibinfo{year}{2017}b.
\newblock \bibinfo{title}{Multilevel contextual 3-d cnns for false positive
  reduction in pulmonary nodule detection}.
\newblock \bibinfo{journal}{IEEE Transactions on Biomedical Engineering}
  \bibinfo{volume}{64}, \bibinfo{pages}{1558--1567}.
\bibitem[{Gidaris et~al.(2018)Gidaris, Singh and
  Komodakis}]{gidaris2018unsupervised}
\bibinfo{author}{Gidaris, S.}, \bibinfo{author}{Singh, P.},
  \bibinfo{author}{Komodakis, N.}, \bibinfo{year}{2018}.
\newblock \bibinfo{title}{Unsupervised representation learning by predicting
  image rotations}.
\newblock \bibinfo{journal}{arXiv preprint arXiv:1803.07728} .
\bibitem[{Girshick(2015)}]{girshick2015fast}
\bibinfo{author}{Girshick, R.}, \bibinfo{year}{2015}.
\newblock \bibinfo{title}{Fast r-cnn}, in: \bibinfo{booktitle}{Proceedings of
  the IEEE international conference on computer vision}, pp.
  \bibinfo{pages}{1440--1448}.
\bibitem[{Gupta et~al.(2018)Gupta, Saar, Martens and
  Moullec}]{gupta2018automatic}
\bibinfo{author}{Gupta, A.}, \bibinfo{author}{Saar, T.},
  \bibinfo{author}{Martens, O.}, \bibinfo{author}{Moullec, Y.L.},
  \bibinfo{year}{2018}.
\newblock \bibinfo{title}{Automatic detection of multisize pulmonary nodules in
  ct images: Large-scale validation of the false-positive reduction step}.
\newblock \bibinfo{journal}{Medical physics} \bibinfo{volume}{45},
  \bibinfo{pages}{1135--1149}.
\bibitem[{Hardie et~al.(2008)Hardie, Rogers, Wilson and
  Rogers}]{hardie2008performance}
\bibinfo{author}{Hardie, R.C.}, \bibinfo{author}{Rogers, S.K.},
  \bibinfo{author}{Wilson, T.}, \bibinfo{author}{Rogers, A.},
  \bibinfo{year}{2008}.
\newblock \bibinfo{title}{Performance analysis of a new computer aided
  detection system for identifying lung nodules on chest radiographs}.
\newblock \bibinfo{journal}{Medical Image Analysis} \bibinfo{volume}{12},
  \bibinfo{pages}{240--258}.
\bibitem[{Hosny et~al.(2018)Hosny, Parmar, Quackenbush, Schwartz and
  Aerts}]{hosny2018artificial}
\bibinfo{author}{Hosny, A.}, \bibinfo{author}{Parmar, C.},
  \bibinfo{author}{Quackenbush, J.}, \bibinfo{author}{Schwartz, L.H.},
  \bibinfo{author}{Aerts, H.J.}, \bibinfo{year}{2018}.
\newblock \bibinfo{title}{Artificial intelligence in radiology}.
\newblock \bibinfo{journal}{Nature Reviews Cancer} \bibinfo{volume}{18},
  \bibinfo{pages}{500}.
\bibitem[{Huang et~al.(2017)Huang, Shan and Vaidya}]{huang2017lung}
\bibinfo{author}{Huang, X.}, \bibinfo{author}{Shan, J.},
  \bibinfo{author}{Vaidya, V.}, \bibinfo{year}{2017}.
\newblock \bibinfo{title}{Lung nodule detection in ct using 3d convolutional
  neural networks}, in: \bibinfo{booktitle}{2017 IEEE 14th International
  Symposium on Biomedical Imaging (ISBI 2017)}, \bibinfo{organization}{IEEE}.
  pp. \bibinfo{pages}{379--383}.
\bibitem[{Jacobs et~al.(2014)Jacobs, van Rikxoort, Twellmann, Scholten,
  de~Jong, Kuhnigk, Oudkerk, de~Koning, Prokop, Schaefer-Prokop
  et~al.}]{jacobs2014automatic}
\bibinfo{author}{Jacobs, C.}, \bibinfo{author}{van Rikxoort, E.M.},
  \bibinfo{author}{Twellmann, T.}, \bibinfo{author}{Scholten, E.T.},
  \bibinfo{author}{de~Jong, P.A.}, \bibinfo{author}{Kuhnigk, J.M.},
  \bibinfo{author}{Oudkerk, M.}, \bibinfo{author}{de~Koning, H.J.},
  \bibinfo{author}{Prokop, M.}, \bibinfo{author}{Schaefer-Prokop, C.}, et~al.,
  \bibinfo{year}{2014}.
\newblock \bibinfo{title}{Automatic detection of subsolid pulmonary nodules in
  thoracic computed tomography images}.
\newblock \bibinfo{journal}{Medical image analysis} \bibinfo{volume}{18},
  \bibinfo{pages}{374--384}.
\bibitem[{Jenuwine et~al.(2018)Jenuwine, Mahesh, Furst and
  Raicu}]{jenuwine2018lung}
\bibinfo{author}{Jenuwine, N.M.}, \bibinfo{author}{Mahesh, S.N.},
  \bibinfo{author}{Furst, J.D.}, \bibinfo{author}{Raicu, D.S.},
  \bibinfo{year}{2018}.
\newblock \bibinfo{title}{Lung nodule detection from ct scans using 3d
  convolutional neural networks without candidate selection}, in:
  \bibinfo{booktitle}{Medical Imaging 2018: Computer-Aided Diagnosis},
  \bibinfo{organization}{International Society for Optics and Photonics}. p.
  \bibinfo{pages}{1057539}.
\bibitem[{Jing and Tian(2018)}]{jing2018self}
\bibinfo{author}{Jing, L.}, \bibinfo{author}{Tian, Y.}, \bibinfo{year}{2018}.
\newblock \bibinfo{title}{Self-supervised spatiotemporal feature learning by
  video geometric transformations}.
\newblock \bibinfo{journal}{arXiv preprint arXiv:1811.11387} .
\bibitem[{Jing and Tian(2019)}]{jing2019self}
\bibinfo{author}{Jing, L.}, \bibinfo{author}{Tian, Y.}, \bibinfo{year}{2019}.
\newblock \bibinfo{title}{Self-supervised visual feature learning with deep
  neural networks: A survey}.
\newblock \bibinfo{journal}{arXiv preprint arXiv:1902.06162} .
\bibitem[{Khosravan and Bagci(2018)}]{khosravan2018s4nd}
\bibinfo{author}{Khosravan, N.}, \bibinfo{author}{Bagci, U.},
  \bibinfo{year}{2018}.
\newblock \bibinfo{title}{S4nd: Single-shot single-scale lung nodule
  detection}, in: \bibinfo{booktitle}{International Conference on Medical Image
  Computing and Computer-Assisted Intervention}, pp. \bibinfo{pages}{794--802}.
\bibitem[{Kim et~al.(2019)Kim, Yoon, Choi and Suk}]{kim2019multi}
\bibinfo{author}{Kim, B.C.}, \bibinfo{author}{Yoon, J.S.},
  \bibinfo{author}{Choi, J.S.}, \bibinfo{author}{Suk, H.I.},
  \bibinfo{year}{2019}.
\newblock \bibinfo{title}{Multi-scale gradual integration cnn for false
  positive reduction in pulmonary nodule detection}.
\newblock \bibinfo{journal}{Neural Networks} .
\bibitem[{Lee et~al.(2017)Lee, Huang, Singh and Yang}]{lee2017unsupervised}
\bibinfo{author}{Lee, H.Y.}, \bibinfo{author}{Huang, J.B.},
  \bibinfo{author}{Singh, M.}, \bibinfo{author}{Yang, M.H.},
  \bibinfo{year}{2017}.
\newblock \bibinfo{title}{Unsupervised representation learning by sorting
  sequences}, in: \bibinfo{booktitle}{Proceedings of the IEEE International
  Conference on Computer Vision}, pp. \bibinfo{pages}{667--676}.
\bibitem[{Liao et~al.(2019)Liao, Liang, Li, Hu and Song}]{liao2019evaluate}
\bibinfo{author}{Liao, F.}, \bibinfo{author}{Liang, M.}, \bibinfo{author}{Li,
  Z.}, \bibinfo{author}{Hu, X.}, \bibinfo{author}{Song, S.},
  \bibinfo{year}{2019}.
\newblock \bibinfo{title}{Evaluate the malignancy of pulmonary nodules using
  the 3-d deep leaky noisy-or network}.
\newblock \bibinfo{journal}{IEEE transactions on neural networks and learning
  systems} .
\bibitem[{Lin et~al.(2017)Lin, Doll{\'a}r, Girshick, He, Hariharan and
  Belongie}]{lin2017feature}
\bibinfo{author}{Lin, T.Y.}, \bibinfo{author}{Doll{\'a}r, P.},
  \bibinfo{author}{Girshick, R.}, \bibinfo{author}{He, K.},
  \bibinfo{author}{Hariharan, B.}, \bibinfo{author}{Belongie, S.},
  \bibinfo{year}{2017}.
\newblock \bibinfo{title}{Feature pyramid networks for object detection}, in:
  \bibinfo{booktitle}{Proceedings of the IEEE Conference on Computer Vision and
  Pattern Recognition}, pp. \bibinfo{pages}{2117--2125}.
\bibitem[{Litjens et~al.(2017)Litjens, Kooi, Bejnordi, Setio, Ciompi,
  Ghafoorian, Van Der~Laak, Van~Ginneken and S{\'a}nchez}]{litjens2017survey}
\bibinfo{author}{Litjens, G.}, \bibinfo{author}{Kooi, T.},
  \bibinfo{author}{Bejnordi, B.E.}, \bibinfo{author}{Setio, A.A.A.},
  \bibinfo{author}{Ciompi, F.}, \bibinfo{author}{Ghafoorian, M.},
  \bibinfo{author}{Van Der~Laak, J.A.}, \bibinfo{author}{Van~Ginneken, B.},
  \bibinfo{author}{S{\'a}nchez, C.I.}, \bibinfo{year}{2017}.
\newblock \bibinfo{title}{A survey on deep learning in medical image analysis}.
\newblock \bibinfo{journal}{Medical image analysis} \bibinfo{volume}{42},
  \bibinfo{pages}{60--88}.
\bibitem[{{Liu} et~al.(2019){Liu}, {Cao}, {Akin} and {Tian}}]{jliu2019miccai}
\bibinfo{author}{{Liu}, J.}, \bibinfo{author}{{Cao}, L.},
  \bibinfo{author}{{Akin}, O.}, \bibinfo{author}{{Tian}, Y.},
  \bibinfo{year}{2019}.
\newblock \bibinfo{title}{{3DFPN-HS$^2$: 3D Feature Pyramid Network Based High
  Sensitivity and Specificity Pulmonary Nodule Detection}}, in:
  \bibinfo{booktitle}{International Conference on Medical Image Computing and
  Computer-Assisted Intervention}.
\bibitem[{Liu et~al.(2016)Liu, Anguelov, Erhan, Szegedy, Reed, Fu and
  Berg}]{liu2016ssd}
\bibinfo{author}{Liu, W.}, \bibinfo{author}{Anguelov, D.},
  \bibinfo{author}{Erhan, D.}, \bibinfo{author}{Szegedy, C.},
  \bibinfo{author}{Reed, S.}, \bibinfo{author}{Fu, C.Y.},
  \bibinfo{author}{Berg, A.C.}, \bibinfo{year}{2016}.
\newblock \bibinfo{title}{Ssd: Single shot multibox detector}, in:
  \bibinfo{booktitle}{European conference on computer vision}, pp.
  \bibinfo{pages}{21--37}.
\bibitem[{Mak et~al.(2019)Mak, Endres, Paik, Sergeev, Aerts, Williams, Lakhani
  and Guinan}]{mak2019use}
\bibinfo{author}{Mak, R.H.}, \bibinfo{author}{Endres, M.G.},
  \bibinfo{author}{Paik, J.H.}, \bibinfo{author}{Sergeev, R.A.},
  \bibinfo{author}{Aerts, H.}, \bibinfo{author}{Williams, C.L.},
  \bibinfo{author}{Lakhani, K.R.}, \bibinfo{author}{Guinan, E.C.},
  \bibinfo{year}{2019}.
\newblock \bibinfo{title}{Use of crowd innovation to develop an artificial
  intelligence--based solution for radiation therapy targeting}.
\newblock \bibinfo{journal}{JAMA oncology} \bibinfo{volume}{5},
  \bibinfo{pages}{654--661}.
\newblock \bibinfo{note}{[dataset]}.
\bibitem[{Matsumoto et~al.(2006)Matsumoto, Kundel, Gee, Gefter and
  Hatabu}]{matsumoto2006pulmonary}
\bibinfo{author}{Matsumoto, S.}, \bibinfo{author}{Kundel, H.L.},
  \bibinfo{author}{Gee, J.C.}, \bibinfo{author}{Gefter, W.B.},
  \bibinfo{author}{Hatabu, H.}, \bibinfo{year}{2006}.
\newblock \bibinfo{title}{Pulmonary nodule detection in ct images with
  quantized convergence index filter}.
\newblock \bibinfo{journal}{Medical Image Analysis} \bibinfo{volume}{10},
  \bibinfo{pages}{343--352}.
\bibitem[{Messay et~al.(2010)Messay, Hardie and Rogers}]{messay2010new}
\bibinfo{author}{Messay, T.}, \bibinfo{author}{Hardie, R.C.},
  \bibinfo{author}{Rogers, S.K.}, \bibinfo{year}{2010}.
\newblock \bibinfo{title}{A new computationally efficient cad system for
  pulmonary nodule detection in ct imagery}.
\newblock \bibinfo{journal}{Medical image analysis} \bibinfo{volume}{14},
  \bibinfo{pages}{390--406}.
\bibitem[{Mundhenk et~al.(2018)Mundhenk, Ho and
  Chen}]{mundhenk2018improvements}
\bibinfo{author}{Mundhenk, T.N.}, \bibinfo{author}{Ho, D.},
  \bibinfo{author}{Chen, B.Y.}, \bibinfo{year}{2018}.
\newblock \bibinfo{title}{Improvements to context based self-supervised
  learning}, in: \bibinfo{booktitle}{Computer Vision and Pattern Recognition
  (CVPR)}.
\bibitem[{Murphy et~al.(2009)Murphy, Van, Schilham, de~Hoop, Gietema and
  Prokop}]{Murphy2009A}
\bibinfo{author}{Murphy, K.}, \bibinfo{author}{Van, G.B.},
  \bibinfo{author}{Schilham, A.M.}, \bibinfo{author}{de~Hoop, B.J.},
  \bibinfo{author}{Gietema, H.A.}, \bibinfo{author}{Prokop, M.},
  \bibinfo{year}{2009}.
\newblock \bibinfo{title}{A large-scale evaluation of automatic pulmonary
  nodule detection in chest ct using local image features and
  k-nearest-neighbour classification}.
\newblock \bibinfo{journal}{Medical Image Analysis} \bibinfo{volume}{13},
  \bibinfo{pages}{757--770}.
\bibitem[{{National Lung Screening Trial
  Research}(2011)}]{national2011national}
\bibinfo{author}{{National Lung Screening Trial Research}, T.},
  \bibinfo{year}{2011}.
\newblock \bibinfo{title}{The national lung screening trial: overview and study
  design}.
\newblock \bibinfo{journal}{Radiology} \bibinfo{volume}{258},
  \bibinfo{pages}{243--253}.
\newblock \bibinfo{note}{[dataset]}.
\bibitem[{Noroozi and Favaro(2016)}]{noroozi2016unsupervised}
\bibinfo{author}{Noroozi, M.}, \bibinfo{author}{Favaro, P.},
  \bibinfo{year}{2016}.
\newblock \bibinfo{title}{Unsupervised learning of visual representations by
  solving jigsaw puzzles}, in: \bibinfo{booktitle}{European Conference on
  Computer Vision}, \bibinfo{organization}{Springer}. pp.
  \bibinfo{pages}{69--84}.
\bibitem[{Ohno et~al.(2017)Ohno, Koyama, Yoshikawa, Kishida, Seki, Takenaka,
  Yui, Miyazaki and Sugimura}]{ohno2017standard}
\bibinfo{author}{Ohno, Y.}, \bibinfo{author}{Koyama, H.},
  \bibinfo{author}{Yoshikawa, T.}, \bibinfo{author}{Kishida, Y.},
  \bibinfo{author}{Seki, S.}, \bibinfo{author}{Takenaka, D.},
  \bibinfo{author}{Yui, M.}, \bibinfo{author}{Miyazaki, M.},
  \bibinfo{author}{Sugimura, K.}, \bibinfo{year}{2017}.
\newblock \bibinfo{title}{Standard-, reduced-, and no-dose thin-section
  radiologic examinations: comparison of capability for nodule detection and
  nodule type assessment in patients suspected of having pulmonary nodules}.
\newblock \bibinfo{journal}{Radiology} \bibinfo{volume}{284},
  \bibinfo{pages}{562--573}.
\bibitem[{Pesce et~al.(2019)Pesce, Withey, Ypsilantis, Bakewell, Goh and
  Montana}]{pesce2019learning}
\bibinfo{author}{Pesce, E.}, \bibinfo{author}{Withey, S.J.},
  \bibinfo{author}{Ypsilantis, P.P.}, \bibinfo{author}{Bakewell, R.},
  \bibinfo{author}{Goh, V.}, \bibinfo{author}{Montana, G.},
  \bibinfo{year}{2019}.
\newblock \bibinfo{title}{Learning to detect chest radiographs containing
  pulmonary lesions using visual attention networks}.
\newblock \bibinfo{journal}{Medical image analysis} \bibinfo{volume}{53},
  \bibinfo{pages}{26--38}.
\bibitem[{Ramachandran et~al.(2018)Ramachandran, George, Skaria and
  Varun}]{ramachandran2018using}
\bibinfo{author}{Ramachandran, S.}, \bibinfo{author}{George, J.},
  \bibinfo{author}{Skaria, S.}, \bibinfo{author}{Varun, V.},
  \bibinfo{year}{2018}.
\newblock \bibinfo{title}{Using yolo based deep learning network for real time
  detection and localization of lung nodules from low dose ct scans}, in:
  \bibinfo{booktitle}{Medical Imaging 2018: Computer-Aided Diagnosis},
  \bibinfo{organization}{International Society for Optics and Photonics}. p.
  \bibinfo{pages}{105751I}.
\bibitem[{Setio et~al.(2016)Setio, Ciompi, Litjens, Gerke, Jacobs, Van,
  Winkler, Naqibullah, Sanchez and Van}]{Setio2016Pulmonary}
\bibinfo{author}{Setio, A.A.}, \bibinfo{author}{Ciompi, F.},
  \bibinfo{author}{Litjens, G.}, \bibinfo{author}{Gerke, P.},
  \bibinfo{author}{Jacobs, C.}, \bibinfo{author}{Van, R.S.},
  \bibinfo{author}{Winkler, W.M.}, \bibinfo{author}{Naqibullah, M.},
  \bibinfo{author}{Sanchez, C.}, \bibinfo{author}{Van, G.B.},
  \bibinfo{year}{2016}.
\newblock \bibinfo{title}{Pulmonary nodule detection in ct images: false
  positive reduction using multi-view convolutional networks}.
\newblock \bibinfo{journal}{IEEE Transactions on Medical Imaging}
  \bibinfo{volume}{35}, \bibinfo{pages}{1160--1169}.
\bibitem[{Shin et~al.(2016)Shin, Roth, Gao, Lu, Xu, Nogues, Yao, Mollura and
  Summers}]{shin2016deep}
\bibinfo{author}{Shin, H.C.}, \bibinfo{author}{Roth, H.R.},
  \bibinfo{author}{Gao, M.}, \bibinfo{author}{Lu, L.}, \bibinfo{author}{Xu,
  Z.}, \bibinfo{author}{Nogues, I.}, \bibinfo{author}{Yao, J.},
  \bibinfo{author}{Mollura, D.}, \bibinfo{author}{Summers, R.M.},
  \bibinfo{year}{2016}.
\newblock \bibinfo{title}{Deep convolutional neural networks for computer-aided
  detection: Cnn architectures, dataset characteristics and transfer learning}.
\newblock \bibinfo{journal}{IEEE Transactions on Medical Imaging}
  \bibinfo{volume}{35}, \bibinfo{pages}{1285}.
\bibitem[{Singh and Davis(2018)}]{singh2018analysis}
\bibinfo{author}{Singh, B.}, \bibinfo{author}{Davis, L.S.},
  \bibinfo{year}{2018}.
\newblock \bibinfo{title}{An analysis of scale invariance in object detection
  snip}, in: \bibinfo{booktitle}{Proceedings of the IEEE Conference on Computer
  Vision and Pattern Recognition}, pp. \bibinfo{pages}{3578--3587}.
\bibitem[{Tan et~al.(2019)Tan, Jing, Huo, Tian and Akin}]{tan2019lgan}
\bibinfo{author}{Tan, J.}, \bibinfo{author}{Jing, L.}, \bibinfo{author}{Huo,
  Y.}, \bibinfo{author}{Tian, Y.}, \bibinfo{author}{Akin, O.},
  \bibinfo{year}{2019}.
\newblock \bibinfo{title}{Lgan: Lung segmentation in ct scans using generative
  adversarial network}.
\newblock \bibinfo{journal}{arXiv preprint arXiv:1901.03473} .
\bibitem[{Van~Ginneken et~al.(2010)Van~Ginneken, Armato~III, de~Hoop, van
  Amelsvoort-van~de Vorst, Duindam, Niemeijer, Murphy, Schilham, Retico,
  Fantacci et~al.}]{van2010comparing}
\bibinfo{author}{Van~Ginneken, B.}, \bibinfo{author}{Armato~III, S.G.},
  \bibinfo{author}{de~Hoop, B.}, \bibinfo{author}{van Amelsvoort-van~de Vorst,
  S.}, \bibinfo{author}{Duindam, T.}, \bibinfo{author}{Niemeijer, M.},
  \bibinfo{author}{Murphy, K.}, \bibinfo{author}{Schilham, A.},
  \bibinfo{author}{Retico, A.}, \bibinfo{author}{Fantacci, M.E.}, et~al.,
  \bibinfo{year}{2010}.
\newblock \bibinfo{title}{Comparing and combining algorithms for computer-aided
  detection of pulmonary nodules in computed tomography scans: the anode09
  study}.
\newblock \bibinfo{journal}{Medical image analysis} \bibinfo{volume}{14},
  \bibinfo{pages}{707--722}.
\bibitem[{Wang et~al.(2018)Wang, Qi, Tang, Zhang, Deng and
  Zhang}]{wang2018automated}
\bibinfo{author}{Wang, B.}, \bibinfo{author}{Qi, G.}, \bibinfo{author}{Tang,
  S.}, \bibinfo{author}{Zhang, L.}, \bibinfo{author}{Deng, L.},
  \bibinfo{author}{Zhang, Y.}, \bibinfo{year}{2018}.
\newblock \bibinfo{title}{Automated pulmonary nodule detection: High
  sensitivity with few candidates}, in: \bibinfo{booktitle}{International
  Conference on Medical Image Computing and Computer-Assisted Intervention},
  \bibinfo{organization}{Springer}. pp. \bibinfo{pages}{759--767}.
\bibitem[{Wang et~al.(2017)Wang, Zhou, Liu, Liu, Gu, Zang, Dong, Gevaert and
  Tian}]{wang2017central}
\bibinfo{author}{Wang, S.}, \bibinfo{author}{Zhou, M.}, \bibinfo{author}{Liu,
  Z.}, \bibinfo{author}{Liu, Z.}, \bibinfo{author}{Gu, D.},
  \bibinfo{author}{Zang, Y.}, \bibinfo{author}{Dong, D.},
  \bibinfo{author}{Gevaert, O.}, \bibinfo{author}{Tian, J.},
  \bibinfo{year}{2017}.
\newblock \bibinfo{title}{Central focused convolutional neural networks:
  Developing a data-driven model for lung nodule segmentation}.
\newblock \bibinfo{journal}{Medical image analysis} \bibinfo{volume}{40},
  \bibinfo{pages}{172--183}.
\bibitem[{Yang et~al.(2012)Yang, Zhang and Tian}]{yang2012recognizing}
\bibinfo{author}{Yang, X.}, \bibinfo{author}{Zhang, C.}, \bibinfo{author}{Tian,
  Y.}, \bibinfo{year}{2012}.
\newblock \bibinfo{title}{Recognizing actions using depth motion maps-based
  histograms of oriented gradients}, in: \bibinfo{booktitle}{Proceedings of the
  20th ACM international conference on Multimedia}, pp.
  \bibinfo{pages}{1057--1060}.
\bibitem[{Zhu et~al.(2018)Zhu, Liu, Fan and Xie}]{zhu2018deeplung}
\bibinfo{author}{Zhu, W.}, \bibinfo{author}{Liu, C.}, \bibinfo{author}{Fan,
  W.}, \bibinfo{author}{Xie, X.}, \bibinfo{year}{2018}.
\newblock \bibinfo{title}{Deeplung: Deep 3d dual path nets for automated
  pulmonary nodule detection and classification}, in: \bibinfo{booktitle}{2018
  IEEE Winter Conference on Applications of Computer Vision (WACV)}, pp.
  \bibinfo{pages}{673--681}.

\end{thebibliography}

\end{document}